\documentclass[superscriptaddress, reprint, amsmath, amssymb, aps, pra, floatfix]{revtex4-2}

\usepackage[utf8]{inputenc}
\usepackage{graphicx}
\usepackage{dcolumn}
\usepackage{bm}
\usepackage[normalem]{ulem} 
\usepackage{mathrsfs}
\usepackage{physics}
\usepackage{xcolor}
\usepackage{amsmath, amssymb, mathtools, amsthm}
\usepackage[colorlinks=true]{hyperref}
\usepackage{comment} 
\usepackage{orcidlink}

\usepackage{xr}

\graphicspath{{Figures/}} 

\definecolor{ve}{RGB}{0,161,22}
\definecolor{dmag}{rgb}{0.6,0.0,0.6}
\definecolor{pink}{rgb}{1,0,0.9}

\begin{abstract}
Entanglement entropy is a fundamental measure of quantum correlations and a key resource underpinning advances in quantum information and many-body physics. We uncover a universal relationship between bipartite entanglement entropy and particle number after the barrier in a one-dimensional Fermi-Hubbard system with an external asymmetric potential. {Decomposing the von Neumann entropy into number entropy~$S_n$ and configurational
entropy~$S_c$, we show that in the barrier-dominated tunneling regime both components
are individually well-defined functions of the post-barrier particle density~$n_A$, even
though~$S_c$ encodes off-diagonal coherences that are not directly accessible from density
measurements alone.} Using Kolmogorov-Arnold Networks---a
novel machine learning architecture---we learn the relationship {for entropy and its components} across a broad range of interaction strengths and barrier heights with high predictive accuracy. Furthermore, we propose a simple analytical binary-entropy-like expression that quantitatively captures the observed correlation for fixed parameters. Our findings open new avenues for characterizing quantum correlations in transport phenomena and provide a powerful framework {for estimating the full von Neumann entropy---including its
configurational component---from a single transport observable.}
\end{abstract}

\begin{document}

\author{Elvira Bilokon\orcidlink{0009-0007-8296-2906}}
\email{ebilokon@tulane.edu}
\affiliation{Department of Physics and Engineering Physics, Tulane University, New Orleans, Louisiana 70118, United States}
\affiliation{Akhiezer Institute for Theoretical Physics, NSC KIPT, Akademichna 1, 61108 Kharkiv, Ukraine}

\author{Valeriia Bilokon\orcidlink{0009-0001-1891-0171}}
\email{vbilokon@tulane.edu}
\affiliation{Department of Physics and Engineering Physics, Tulane University, New Orleans, Louisiana 70118, United States}
\affiliation{Akhiezer Institute for Theoretical Physics, NSC KIPT, Akademichna 1, 61108 Kharkiv, Ukraine}

\author{Abhijit Sen \orcidlink{0000-0003-2783-1763}}
\email{abhijit913@gmail.com}
\affiliation{Department of Physics and Engineering Physics, Tulane University,  New Orleans, Louisiana 70118, United States}

\author{Mohammed Th. Hassan\orcidlink{0000-0002-9669-534X}}
\email{mohammedhassan@arizona.edu}
\affiliation{Department of Physics, University of Arizona, Tucson, AZ, USA}
\affiliation{James C. Wyant College of Optical Sciences, University of Arizona, Tucson, AZ, United States}

\author{Andrii G. Sotnikov\orcidlink{0000-0002-3632-4790}}
\email{a\_sotnikov@kipt.kharkov.ua}
\affiliation{Akhiezer Institute for Theoretical Physics, NSC KIPT, Akademichna 1, 61108 Kharkiv, Ukraine}
\affiliation{Karazin Kharkiv National University, Svobody Square 4, 61022 Kharkiv, Ukraine}

\author{Denys I. Bondar \orcidlink{0000-0002-3626-4804}}
\email{dbondar@tulane.edu}
\affiliation{Department of Physics and Engineering Physics, Tulane University, New Orleans, Louisiana 70118, United States}

\title{{Predicting Entanglement Entropy from Particle Tunneling of Interacting Fermions Using Kolmogorov-Arnold Networks}}

\maketitle

Quantifying entanglement entropy in many-body quantum systems remains one of the most challenging yet fundamental tasks in quantum physics~\cite{Amico2008, Eisert2010}. The von Neumann entropy, which captures the quantum correlations between spatially separated regions, requires full knowledge of the many-body wavefunction—a formidable computational and experimental challenge that scales exponentially with system size. Direct measurement of entanglement entropy demands sophisticated quantum state tomography techniques~\cite{James2001} that become prohibitively complex for large systems. While alternative approaches~\cite{Islam2015, Brydges2019, Pichler2013, Elben2018, Hartman2018, Lin2024}, including machine learning-based techniques~\cite{Koutny2023, Huang2025}, have provided valuable insights, they either offer only partial characterization or remain computationally intensive for large systems, which limits the ability to probe quantum correlations in realistic many-body settings~\cite{Brydges2019, Elbein2020}.

In stark contrast, particle transport measurements have become remarkably accessible in modern quantum simulation platforms. Ultracold atoms trapped in optical lattices~\cite{Folling2007, Trotzky2008, Spielman2007, Zwierlein2004, Shin2006} now allow unprecedented control and monitoring of single-particle dynamics with site-resolved precision~\cite{Miranda2015,Yamamoto2016, Young2022, Parsons2016, Cheuk2016}. These experimental advances enable direct observation of particle tunneling, density distributions, and transport phenomena in real-time, providing a natural window into the underlying quantum dynamics~\cite{Brown2019, Nichols2019}. The ability to engineer controlled barriers, tune interaction strengths, and track individual particles makes ultracold atoms ideal testbeds for exploring many-body systems.

These developments breathe new life into foundational quantum phenomena, bringing well-established concepts like tunneling into dialogue with entropy---a modern theoretical tool. Quantum tunneling stands among the earliest and most fundamental discoveries in quantum physics, dating back to the pioneering work of the 1920s and serving as a cornerstone of our understanding of quantum mechanics~\cite{Gurney1928, Gamow1928}. In contrast, entanglement entropy has emerged as a central concept only in recent decades, driven by advances in quantum information theory and the growing recognition of its role in characterizing quantum correlations~\cite{Bennett1996, Nielsen2010}. 
In this broader context, entanglement entropy has previously been related to measurable quantities in various settings:
particle-number fluctuations in equilibrium systems~\cite{Klich2006}, current noise and full counting statistics in nonequilibrium quantum point contacts with noninteracting fermions~\cite{Klich2009}, and density correlations following quantum quenches in one-dimensional integrable models~\cite{Maestro2022}.
These connections, in turn, rest on earlier results establishing universal scaling of entanglement entropy in conformal field theories and its formulation in terms of correlation matrices for Gaussian many-body states~\cite{Calabrese2004, Calabrese2006, Peschel2003}.
However, analytical methods remain unavailable for generic interacting fermions.
Building on these foundations, we combine the time-tested physics of barrier tunneling with modern entanglement measures to reveal functional relationships in interacting systems, providing a pathway to probe quantum correlations through readily accessible transport observables.

In this Letter,
we uncover a remarkable relationship that bridges this experimental divide: the bipartite entanglement entropy $S(t)$ exhibits a functional dependence on particle density $n(t)$ in the one-dimensional Fermi-Hubbard model, i.e., $S(t) = f\big(n(t)\big)$. Crucially, this relationship depends only on the particle density, not on time itself $\Big[S(t) \neq f\big(t,n(t)\big)\Big]$, and reveals a connection independent of the specific dynamical trajectory. {By decomposing the entanglement entropy into its number and configurational components, we show that not only $S_n$ but also $S_c$---which generally encodes information inaccessible from density measurements---is functionally determined by the post-barrier particle density in the barrier-dominated tunneling regime.} By employing Kolmogorov-Arnold Networks~(KANs)~\cite{Liu2025}, we uncover a deeper structure in the entropy-transport correlation, revealing a multivariate dependence $S(t) = f\big(U, h, n(t)\big)$, where the entanglement {and its components} can be learned as a function of interaction strength $U$, barrier height $h$, and particle density $n(t)$ across diverse parameter regimes with high accuracy.
Remarkably, we find that for fixed parameters of the interacting system, this complex relationship reduces to an analytical form resembling binary entropy {for all entropy components}, providing an unexpected connection between quantum information theory and transport phenomena. 
In this way, a synthesis of old and new quantum concepts establishes an efficient framework for inferring quantum correlations{---including the generally inaccessible configurational entropy---}from readily measurable transport observables, opening new avenues for characterizing entanglement in quantum many-body systems without the need for complex state reconstruction protocols.

\begin{figure}
    \centering
    \includegraphics[width=0.75\linewidth]{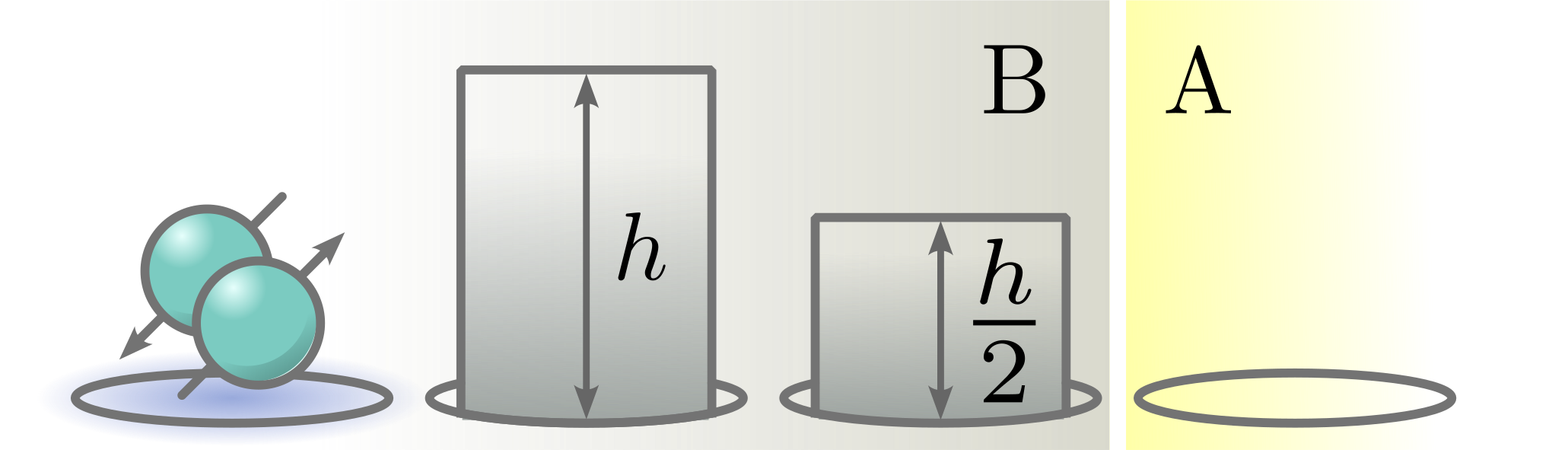}
    \caption{Graphical representation  of the model setup. Initially, $N$ fermions are localized on the very left side. The external potential barrier with the maximum height $h$ acts on the two central sites, dividing the system into subsystem~$B$ (containing the sites with the barrier sites and before it, highlighted in gray) and subsystem $A$ (post-barrier region, highlighted in yellow).}
    \label{fig:model}
\end{figure}

\textit{Model and Methods---}We study quantum correlations and particle transport in a one-dimensional lattice of the size~$L$ with open boundary conditions. The system is described by the single-band Fermi-Hubbard Hamiltonian with an external {field forming an additional} potential barrier:
\begin{equation}
    \hat{\mathcal{H}} = \hat{\mathcal{H}}_{\rm hop} + \hat{\mathcal{H}}_{\rm int} + \hat{\mathcal{H}}_{\rm barrier},
\end{equation}
where 
\begin{align}
    &\hat{\mathcal{H}}_{\rm hop} = -J \sum_{j=1, \sigma = \uparrow, \downarrow}^{L-1} \left( \hat{c}_{j, \sigma}^{\dagger} \hat{c}^{}_{j+1, \sigma} + \hat{c}_{j+1, \sigma}^{\dagger} \hat{c}^{}_{j, \sigma} \right), \\ \nonumber
    &\hat{\mathcal{H}}_{\rm int} = U \sum_{j=1}^L \hat{n}_{j, \uparrow} \hat{n}_{j, \downarrow}, \quad
    \hat{\mathcal{H}}_{\rm barrier} = \sum_{j= L/2}^{L/2+1} V^{\rm ex}_j\hat{n}_{j}.
\end{align}
Here, $\hat{c}_{j,\sigma}^\dagger$ ($\hat{c}_{j,\sigma}$) is the fermionic creation (annihilation) operator for a particle with spin $\sigma \in \{ \uparrow, \downarrow \}$ at site $j$, and $\hat{n}_{j,\sigma} = \hat{c}_{j,\sigma}^\dagger \hat{c}_{j,\sigma}$ is the corresponding number operator, $\hat{n}_j=\hat{n}_{j, \uparrow}+\hat{n}_{j, \downarrow}$. The parameters $J>0$, $U$, and $V^{\rm ex}_j$ represent the hopping amplitude, the on-site repulsive interaction, and the site-dependent barrier height, respectively. The external potential acts only on the two central lattice sites, creating an asymmetric
barrier that divides the system. The site-dependent barrier heights are defined as:
\begin{equation}\label{eq:potential}
    V^{\rm ex}_j = \begin{cases}
    h, & j = L/2 \\
    h/2, & j = L/2+1
\end{cases}
\end{equation}
where $h$ is the maximum barrier height, creating an asymmetric potential configuration as illustrated in Fig.~\ref{fig:model}.

We partition the lattice into subsystem $A$ (all sites after the barrier) and subsystem $B$ (the remaining sites, including those with the barrier and all sites before it), as illustrated in Fig.~\ref{fig:model}. At the initial moment of time ($t=0$) we place $N$ particles on the left side before the barrier. 

The key transport observable is the expectation value of the number operator $\hat{n}_{A} =\sum_{j\in A} \hat{n}_j$, where $j$ runs over all sites of subsystem $A$. To characterize the quantum correlations in the system described by the state $|\psi\rangle$, we calculate the von Neumann entanglement entropy {$S_A$}:
\begin{equation}
    {S_A} = -{\rm Tr} (\rho_{A} \ln\rho_{A}),
\end{equation}
where $\rho_{A} = {\rm Tr}_B |\psi\rangle\langle\psi|$ is the reduced density matrix of subsystem $A$. Because the total particle number $N$ is conserved, $\rho_A$ is block-diagonal in sectors of fixed particle number $\tilde{N}$ within subsystem $A$.
Each sector has probability $p_{\tilde{N}} = \mathrm{Tr}(\Pi_{\tilde{N}} \rho_A)$, where $\Pi_{\tilde{N}}$ projects onto states with exactly $\tilde{N}$ particles in $A$. This structure yields a natural decomposition of the entanglement entropy into two contributions~\cite{Lukin2019}:
\begin{equation}
  S_A = S_n + S_c,
  \label{eq:decomposition}
\end{equation}
where the number entropy
\begin{equation}
  S_n = -\sum_{\tilde{N}} p_{\tilde{N}} \ln p_{\tilde{N}}
  \label{eq:Sn}
\end{equation}
captures classical fluctuations in the particle number across the partition, and the configurational entropy
\begin{equation}
  S_c = \sum_{\tilde{N}} p_{\tilde{N}} \, S(\rho_A^{({\tilde{N}})}/p_{\tilde{N}})
  \label{eq:Sc}
\end{equation}
quantifies quantum correlations within each
fixed-particle-number sector, with $\rho_A^{({\tilde{N}})} = \Pi_{\tilde{N}} \rho_A \Pi_{\tilde{N}}$. While $S_{\tilde{N}}$ is experimentally accessible through site-resolved density measurements, the probabilities $p_{\tilde{N}}$ can be extracted from repeated snapshots of atom number in the subsystem~\cite{Lukin2019}. The configurational entropy $S_c$ encodes off-diagonal coherences that are not directly observable from density data alone.

Time evolution calculations are performed using the QuSpin package~\cite{Weinberg2017, Weinberg2019}. We evolve the initial state $\ket{\psi(t=0)}$ over a sufficiently long time interval ($t\sim5000/J$ in units of $\hbar=1$) to resolve the particle tunneling dynamics. Our simulation timescales are consistent with the experimental capabilities of current cold-atom lattice setups \cite{Brown2019, Nichols2019, Parsons2016}. To extract the correlations between $S_A$ and $\langle \hat{n}_{A}\rangle$, we employ Kolmogorov-Arnold Networks~\cite{Liu2025}, a recently developed and effective neural network architecture based on the Kolmogorov-Arnold representation theorem~\cite{Kolmogorov1957, Arnold2009}. In KANs, learnable activation functions are placed on network edges and parameterized as weighted combinations of basis functions, typically implemented using B-splines.

\textit{Results---}We study the bipartite entanglement entropy~$S_A(t)$, which characterizes quantum correlations between post-barrier sites and the remaining system. To reveal the underlying relationship between entanglement and particle transport, we analyze the dependence $S_A(n_{A})$, where $n_{A} = \langle\hat{n}_{A}\rangle(t)$. This directly correlates entanglement with particle occupation probability after the barrier.

\begin{figure}
    \centering
    \includegraphics[width=\linewidth]{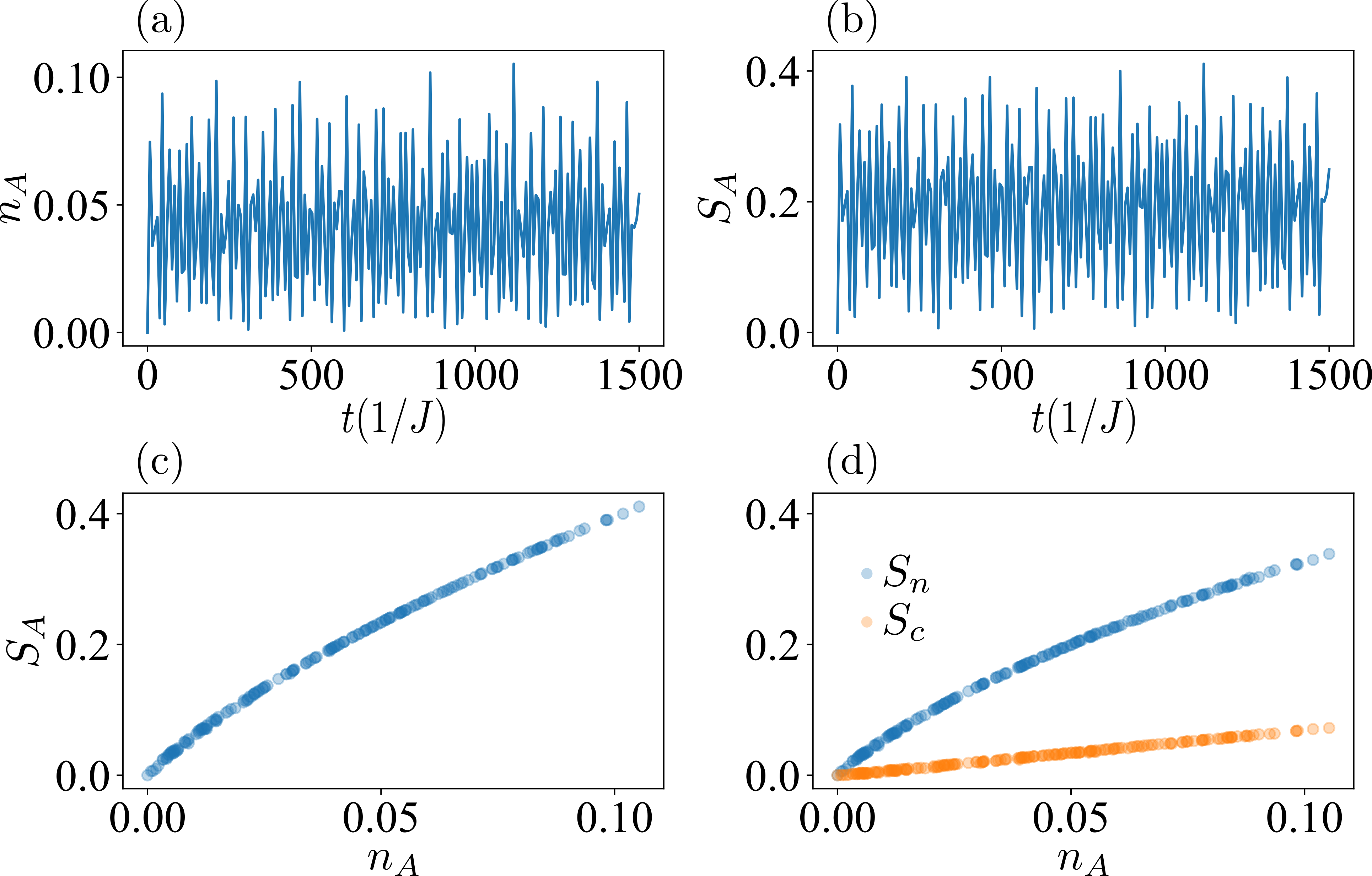}
    \caption{{Entropy--density relationship and its decomposition. 
    (a) Particle density $n_A$ as a function of time. 
    (b) Entanglement entropy $S_A$ as a function of time. 
    (c) Von Neumann entanglement entropy $S_A$ plotted against the post-barrier particle density $n_A$. 
    (d) Corresponding number and configurational entropy contributions, $S_n$ and $S_c$, plotted against $n_A$. 
    Each point in panels (c,d) represents one time step of the evolution over $tJ \in [0,1500]$. 
    Parameters are $L=4$, $N=2$, $U=2J$, and $h=5J$.}}
    \label{fig:ent_vs_dens}
\end{figure}

{Figure~\ref{fig:ent_vs_dens}(a) demonstrates the relation between $S_A$ and ${n}_{A}$ for $L=4$, $U=2J$, and $h=5J$.} While individually ${n}_{A}$ [Fig.~\ref{fig:ent_vs_dens}(c)] and $S$ [Fig.~\ref{fig:ent_vs_dens}(d)] exhibit complex temporal fluctuations, the plot $S_A(n_{A})$ reveals a clear dependence. At $n_A\approx0$, corresponding to early times when particles have not yet tunneled through the barrier, $S_A$ remains minimal, reflecting weak correlations between subsystems $A$ and $B$. As particles tunnel through the barrier, $S_A$ increases exhibiting a characteristic functional dependence on particle occupation changes.

{To understand the origin of this relationship, we decompose von Neumann entropy into number entropy $S_n$ and configurational entropy~$S_c$ defined in Eqs.~\eqref{eq:Sn}-\eqref{eq:Sc}. Figure~\ref{fig:ent_vs_dens}(b) shows that both components $S_n(n_A)$ and $S_c(n_A)$ separately demonstrate a clean relationship with the post-barrier density. The configurational entropy~$S_c$ constitutes a non-negligible fraction of the total entropy, yet it is itself functionally determined by $n_A$ in the barrier-dominated tunneling regime.  }

\textit{Utilizing KAN---}While $S_A(n_{A})$ reveals clear correlations for fixed parameters, understanding how interaction strength and barrier height influence this relationship requires exploring the multivariate dependence $S_A(U, h, n_{A})$. {To this end, we employ Kolmogorov-Arnold Networks (for architecture and training details, see End Matter).}

\begin{figure}
    \centering
    \includegraphics[width=\linewidth]{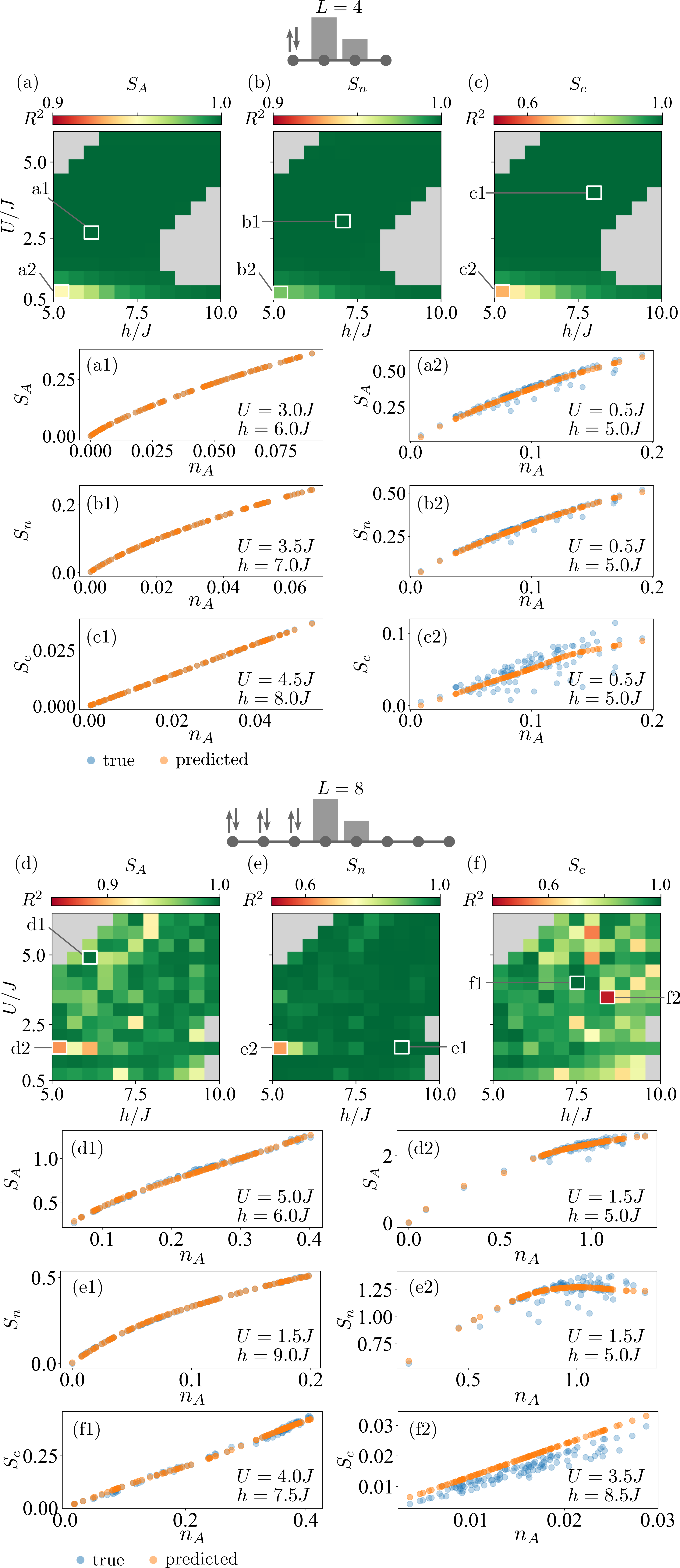}
    \caption{{KAN prediction accuracy for the entropy--density relationship across the $(U, h)$ parameter space, shown for the worst-performing cross-validation fold. $R^2$ heatmaps for $S_{A}$, $S_n$, and $S_c$ for $L = 4$, $N = 2$ [(a)--(c)] and $L = 8$, $N = 6$ [(d)--(f)]. Gray cells indicate excluded parameter pairs where $U \geq h$ or $\max n_A(t) < 0.01$. (a1--c2) and (d1--f2): comparison of true (blue) and KAN-predicted (orange) values of $S_{A}(n_A)$, $S_n(n_A)$, and $S_c(n_A)$ for the best and worst $R^2$ parameter pairs. Time evolution is performed over $tJ \in [0, 1000]$. The lattice diagrams show the system setup with fermions initially localized on the left and external potential barriers (gray boxes) acting on two central sites.}}
    \label{fig:kan}
\end{figure}

{The choice of parameter ranges is dictated by the
tunneling physics. We restrict $h \in [5J, 10J]$ to ensure that transport occurs via tunneling rather than over-barrier transmission. Simultaneously, the on-site interaction must remain below the barrier,
$U < h$, to prevent interaction-assisted over-barrier processes; we scan $U \in [0.5J, 10J]$ and exclude all pairs with $U \geq h$. We further filter the dataset by requiring that the maximum value of $n_A(t) > 0.01$ over the full time evolution, removing configurations in which the barrier suppresses tunneling. In the $R^2$ heatmaps of
Fig.~\ref{fig:kan}, these excluded configurations appear as gray cells.}

For each $(U, h$) pair, we perform time evolution over $tJ\in[0,1000]$ and collect data $S_A(n_{A})$. Combining datasets across multiple $(U, h)$ values creates a unified training set where $U$, $h$, and $n_{A}$ serve as independent inputs for learning $S_A(U, h, n_{A})$. {Robustness is verified through 5-fold cross-validation with results reported in the Supplemental Material.}

    We train separate KAN models for $S_A$ and its components, $S_n$ and $S_c$. Figure~\ref{fig:kan} summarizes the results for the worst-performing cross-validation fold. For both system sizes, the number entropy $S_n$ remains a well-defined function of $n_A$ across the full parameter space. KAN achieves $R^2>0.9$ for almost all included $(U, h)$ pairs, consistent with $S_n$ being determined by the particle-number distribution alone. 

    The configurational entropy behaves differently as the system grows. For $L = 4$, $N = 2$, $S_c$ stays small and retains a clean functional dependence on $n_A$, yielding $R^2 > 0.9$ across all parameter pairs. For the larger system ($L = 8$, $N = 6$), $S_c$ grows in magnitude and exhibits a more scattered dependence on
$n_A$: as the number of post-barrier sites increases, tunneled particles have more configurations in which to distribute, and the within-sector correlations encoded by $S_c$ become sensitive to details beyond the mean density. This is reflected in the $S_c$ heatmap, where
$R^2$ degrades notably at low $h/U$ ratios. However, despite this increased scatter, the values of $S_c$ remain bounded, and KAN still captures the overall magnitude and trend of $S_c(n_A)$. Consequently, the total entanglement entropy $S_{A}$ is predicted with high accuracy throughout the barrier-dominated regime.

\textit{Analytical Description of the Entropy-Density Relation---}The KAN analysis confirms the existence of a multivariate relationship but fails to build a symbolic expression using pruning~\cite{Liu2025}. To address this limitation, we pursued an alternative approach based on physical intuition.
 We hypothesized that the entropy-density relationship should reflect the nature of quantum entanglement in bipartite systems. Specifically, since our system involves a binary choice for particle localization (before or after the barrier), we expected the entanglement entropy to follow a form similar to binary entropy. This intuition led us to propose that the {$S_{\alpha}(n_{A})$ relationship can be accurately described by: 
 \begin{equation}\label{eq:analytical_approx}
     S_{\alpha}(n_{A}) = c_1^{\alpha} p\ln p + c_2^{\alpha}\left(1 - p\right)\ln\left(1 - p\right), \quad p = \frac{n_{A}}{N},
 \end{equation}
where $\alpha\in\{A, n, c\}$ labels the entropy component and $c_1^{\alpha}$, $c_2^{\alpha}$ are fitting coefficients that depend on the interaction strength $U$ and barrier height $h$.} These coefficients exhibit complex parameter dependence without simple analytical scaling relations, especially for larger system sizes. The first term captures the contribution from particles that have successfully tunneled into subsystem~$A$ (the post-barrier region). Meanwhile, the second term describes the contribution from the particles that remain localized in subsystem~$B$ (the pre-barrier region, including the barrier sites themselves). {For the number entropy, this structure follows directly from the particle-number distribution $p_{\tilde{N}}$; for $S_c$ and $S_{A}$, it serves as an effective parametrization of the more complex within-sector correlations.}

\begin{figure}
    \centering
    \includegraphics[width=\linewidth]{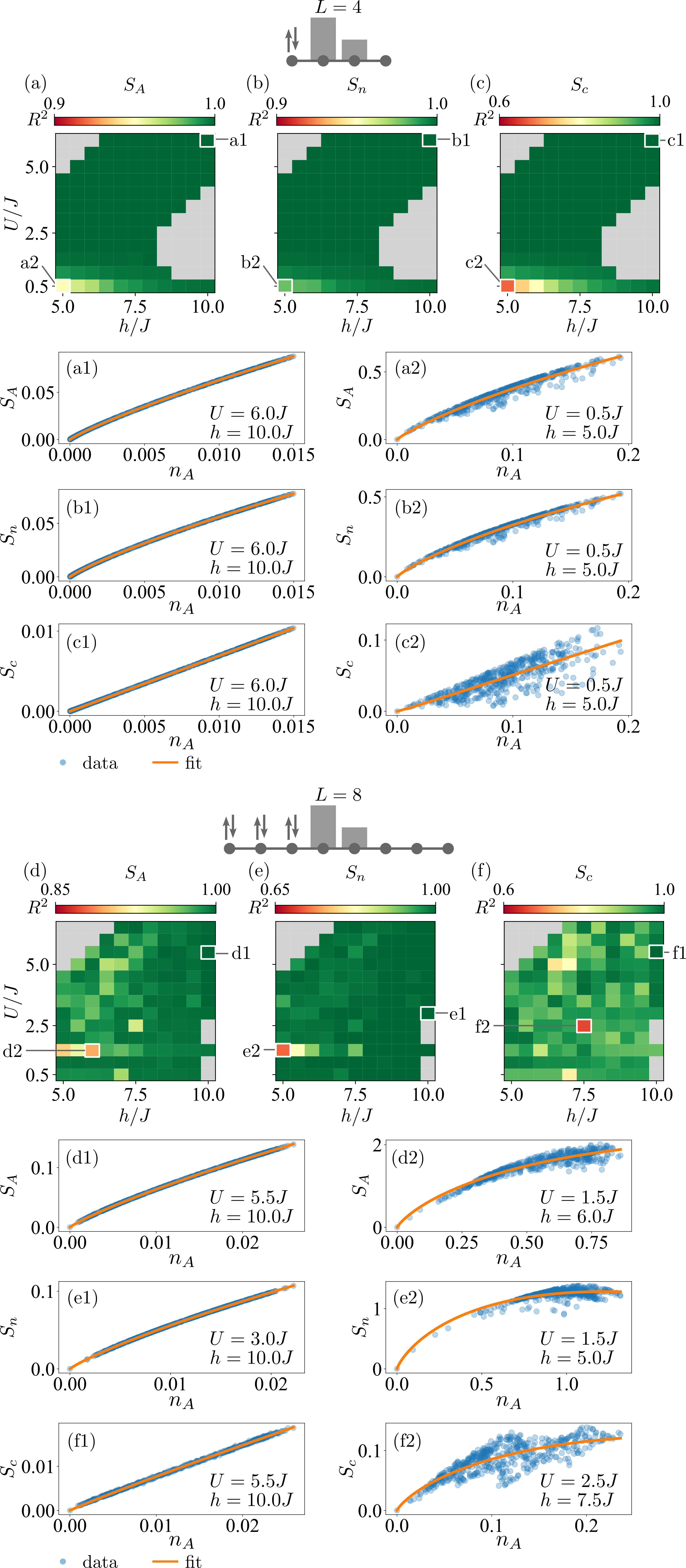}
    \caption{{Validation of the analytical approximation for the entropy-density relationship. $R^2$ heatmaps for $S_{A}$, $S_n$, and $S_c$ for $L = 4$, $N = 2$ [(a)--(c)] and $L = 8$, $N = 6$ [(d)--(f)]. Gray cells indicate excluded parameter pairs where $U \geq h$ or $\max n_A(t) < 0.01$. (a1--c2) and (d1--f2): comparison of true (blue scatter) values and analytical fit (orange curve) of $S_{A}(n_A)$, $S_n(n_A)$, and $S_c(n_A)$ for the best and worst $R^2$ parameter pairs. Time evolution is performed over $tJ \in [0, 1000]$. The lattice diagrams show the system setup with fermions initially localized on the left and external potential barriers (gray boxes) acting on two central sites.}}
    \label{fig:coeffs}
\end{figure}

{We perform least-squares fitting of
Eq.~(\ref{eq:analytical_approx}) independently for $S_{A}$,
$S_n$, and $S_c$ at each $(U,h)$ pair, using the same parameter ranges and filtering criteria as in the KAN
analysis. Figure~\ref{fig:coeffs} displays the
resulting $R^2$ heatmaps and representative $S_{\alpha}(n_A)$ plots for the best (worst) $R^2$ values for both system sizes. For $L = 4$, $N = 2$, the analytical form achieves $R^2 > 0.9$ for $S_A$ and $S_n$ entropies across the entire parameter
space, mirroring the KAN results. The $S_c$ fit quality is comparable, with the lowest $R^2$ value corresponding to the weakly interacting system of $U=0.5 J$. For $L = 8$, $N = 6$,
the fit quality for $S_{A}$ and $S_n$ remains high, while
$S_c$ again shows reduced $R^2$ at low $h/U$---consistent
with the increased scatter in $S_c(n_A)$ identified in the
KAN analysis. The close agreement between KAN and the
analytical fit confirms that the binary entropy structure
of Eq.~(\ref{eq:analytical_approx}) captures the essential
functional dependence.} 

\textit{Discussion---}{A central finding of this work is that in the barrier-dominated tunneling regime, not only the number entropy $S_n$ but also the configurational entropy $S_c$ can be predicted from the post-barrier particle density. This is noteworthy because $S_c$ encodes off-diagonal coherences within each particle-number sector of $\rho_A$---information that is, in general, not accessible from density measurements alone~\cite{Lukin2019}. In the regime studied here, however, the constrained tunneling dynamics couple the growth of $S_c$ to particle transfer, so that $S_c(n_A)$ remains a well-defined function of $n_A$ for small systems and a bounded, predictable quantity even as the system size grows. Combined with the analytical binary entropy structure of Eq.~(\ref{eq:analytical_approx}), this provides a practical route to estimating the full von Neumann entropy---including its configurational component---from a single transport observable.}

All detailed calculations and KAN analysis were performed for systems with maximum $L=8$ lattice sites. However, it is crucial to establish that the entropy-density relationship persists in larger systems. Our preliminary calculations demonstrate the presence of the functional dependence for systems as large as $L=20$ sites, indicating that the observed correlations are not artifacts of finite-size effects but represent a fundamental feature of quantum transport in interacting lattice systems. The scalability of this relationship suggests potential applications to realistic experimental setups. 

The necessary condition for observing the functional dependence is that initially the post-barrier region must be empty, with all particles localized on the opposite side of the potential barrier. This configuration ensures that the entanglement dynamics arise solely from quantum correlations generated during the tunneling process, rather than from pre-existing entanglement.

While our analysis primarily employs an asymmetric triangular barrier, the functional relationship $S_A(n_A)$ is not specific to this particular barrier geometry. We have verified that the correlation persists for rectangular (symmetric) barriers as well as different triangular profiles. The asymmetric triangular barrier was chosen because it reveals new fundamental transport phenomena that are absent in systems with symmetric barriers~\cite{Bilokon2025}. The universal feature is the binary entropy structure itself, not the specific potential shape. 
Our findings link quantum correlations and transport, offering a powerful framework for predicting entanglement dynamics in interacting systems, which potentially can be measured by current tunneling mechanism~\cite{Sennary2025}.


\textit{Acknowledgments}
{We thank the anonymous Referee for suggesting the study of configurational entropy, which substantially improved the manuscript.}
E.B. was supported by the National Science Foundation (NSF) IMPRESS-U Grant No.~2403609. A.S. and D.I.B. was supported by Army Research Office (ARO) (grant W911NF-23-1-0288; program manager Dr.~James Joseph). A.G.S. acknowledges support by the National Research Foundation of Ukraine, project No.~2023.03/0073. The views and conclusions contained in this document are those of the authors and should not be interpreted as representing the official policies, either expressed or implied, of ARO, NSF, or the U.S. Government. The U.S. Government is authorized to reproduce and distribute reprints for Government purposes notwithstanding any copyright notation herein. 

\bibliography{main}

\section*{End Matter}

\textit{KAN Architecture and Training.}---We implement KANs with 3 input nodes
(corresponding to $U$, $h$, and $n_A$) and 1 output node ($S_\alpha$).
For the smaller system ($L = 4$, $N = 2$), we use 3 hidden nodes
in a single hidden layer, resulting in a $[3, 3, 1]$ architecture
with a grid size of 10. For the larger system ($L = 8$, $N = 6$),
we employ a two-hidden-layer architecture $[3, 6, 3, 1]$ with a
grid size of 20. The learnable activation functions on each edge
are parameterized using cubic B-splines. Training is performed
for 50 epochs using the L-BFGS optimizer with a learning rate
of 0.01. All model parameters are summarized in Table~\ref{tab:kan_params}.

\begin{table}[h]
\centering
\begin{tabular}{| c | c | c | c | c | c |}
\hline
$L$ & Architecture & Spline order & Grid size & Learning rate & Loss \\
\hline
4 & $[3, 3, 1]$     & 3 & 10 & 0.01 & $10^{-3}$ \\
8 & $[3, 6, 3, 1]$  & 3 & 20 & 0.01 & $10^{-3}$ \\
\hline
\end{tabular}
\caption{KAN parameters used to model the $S_\alpha(U, h, n_A)$
functional relationship.}
\label{tab:kan_params}
\end{table}

To ensure the robustness of the model, we employ 5-fold
cross-validation on the combined dataset for each $S_\alpha$ individually. The data is partitioned such that each fold contains representative samples from all $(U, h)$ pairs. Table~\ref{tab:cross_val} reports the $R^2$ values for all folds.
Across all five validation folds, the KAN achieves consistent performance confirming robust predictive accuracy across different system configurations.

\begin{table}[h]
\centering
\begin{tabular}{| c | c | c | c |}
\hline
\multicolumn{4}{|c|}{$L=4$} \\
\hline
Fold & $S_A$ & $S_n$ & $S_c$ \\
\hline
1 & 0.9982 & 0.9992 & 0.9837 \\
2 & 0.9984 & 0.9992 & 0.9840 \\
3 & 0.9981 & 0.9991 & 0.9811 \\
4 & 0.9981 & 0.9991 & 0.9833 \\
5 & 0.9979 & 0.9989 & 0.9809 \\
\hline
Mean & $0.9981 \pm 0.0001$ & $0.9991 \pm 0.0001$ & $0.9826 \pm 0.0013$ \\
\hline \hline
\multicolumn{4}{|c|}{$L=8$} \\
\hline
Fold & $S_A$ & $S_n$ & $S_c$ \\
\hline
1 & 0.9985 & 0.9986 & 0.9964 \\
2 & 0.9986 & 0.9989 & 0.9965 \\
3 & 0.9985 & 0.9988 & 0.9962 \\
4 & 0.9985 & 0.9987 & 0.9963 \\
5 & 0.9986 & 0.9989 & 0.9963 \\
\hline
Mean & $0.9986 \pm 0.0001$ & $0.9988 \pm 0.0001$ & $0.9963 \pm 0.0001$ \\
\hline
\end{tabular}
\caption{Cross-validation results for KAN model performance.
$R^2$ values across 5 folds for lattice sizes $L = 4$, $N = 2$
and $L = 8$, $N = 6$.}
\label{tab:cross_val}
\end{table}

\end{document}